%% file: final.tex
\documentclass[usenatbib]{mn2e}
\usepackage{setspace}
\input{psfig}
\input{macros_yun}

\begin{document}


\title[On Halo Formation Times and Assembly Bias]
      {On Halo Formation Times and Assembly Bias}
\author[Li, Mo \& Gao]
       {Yun~Li$^1$\thanks{E-mail: liyun@astro.umass.edu}, H.J.~Mo$^1$, L. Gao$^2$\\
        $^1$Department of Astronomy, University of Massachusetts, 
            MA 01003, USA\\
	$^2$Institute of Computational Cosmology, Department of Physics,
	   University of Durham, \\Science Laboratories, South Road, Durham DH1
	   3LE, UK \\
	}


\date{}

\pagerange{\pageref{firstpage}--\pageref{lastpage}}
\pubyear{2000}

\maketitle

\label{firstpage}


\begin{abstract}

In this paper we use the ``Millennium Simulation'' to 
re-examine the mass assembly history of dark matter halos and the age 
dependence of halo clustering. We use eight different definitions 
of halo formation times to characterize the different aspects of the 
assembly history of a dark matter halo. We find that these formation
times have different dependence on halo mass. While some formation
times characterize well the hierarchical nature of halo formation, 
in the sense that more massive halos have later formation, the trend 
is reversed for other definitions of the formation time. 
In particular, the formation times that are likely to be related 
to the formation of galaxies in dark halos show strong trends 
of ``down-sizing'', in that lower-mass halos form later. 
We also investigate how the correlation amplitude of dark matter 
halos depends on the different formation times. We find that 
this dependence is quite strong for some definitions 
of formation time but weak or absent for other definitions.
In particular, the correlation amplitude of halos of a given 
mass is almost independent of their last major merger time. 
For the definitions that are expected to be more related to the 
formation of galaxies in dark halos, a significant assembly bias 
is found only for halos less massive than $\Mstar$.  
We discuss our results in connection to the hierarchical assembly 
of dark matter halos, the ``archaeological down-sizing''
observed in the galaxy population, and the observed 
color-dependence of the clustering strength of galaxy 
groups and clusters.    
\end{abstract}


\section{Introduction}
\label{sec:intro}
The cold dark matter (CDM) scenario has been proven a very successful model for
structure formation. In this framework, most mass in the cosmic density field 
ends up in virialized objects, called dark matter halos, and luminous objects,
such as galaxies, are supposed to form and evolve in 
such halos~\citep{WR78}. Clearly, the first step in understanding the formation of 
galaxies is to understand the formation of the dark halo population.
Using both $N$-body simulations and semi-analytical methods, 
many important results have been obtained regarding the properties
of dark matter halos in the current CDM paradigm of structure formation, 
including halo mass function ~\citep[e.g.][]{Bond91,LC93,ST99,SMT01,War06}, 
density and sub-halo profile~\citep[e.g.][]{NFW97,Bul01b,ENS,Gao04,Lu06}, 
angular momentum property~\citep[e.g.][]{BE87,CL96,Bul01a,Chen02},
and clustering property ~\citep[e.g.][]{Mo96,Jing98,Lem99,ST99,SMT01}. 
These results have been playing an important role in our understanding 
of galaxy formation. 

 One important property of a dark halo is its formation history.
In the previous studies, this formation history is usually 
characterized by a single parameter which is the time when a halo 
has acquired half of its final halo mass 
~\citep[e.g.][]{LC93,Lem99,Frank02,Gao05,Wec06}. 
This definition of halo formation time is useful because it 
indicates when the main body of a halo is assembled. 
However, it is unclear if such definition is closely related to 
how galaxies form in a halo. For example, \citet[][]{Frank03} 
and \citet[][]{Yang03} both found that dark halos with masses 
around $10^{11.5}\msunh$ have the lowest mass-to-light ratio,
which suggests that star formation is the most efficient in halos 
with a fixed mass around $10^{11-12}\msunh$. Thus, for halos 
with masses much larger than this mass, the half-mass assembly 
time may have little to do with how galaxies may have formed 
in such halos. Based on the half mass formation time, 
more massive halos are expected to form later due to 
hierarchical clustering. This is in contrast with the recent 
observations that the stellar population in more massive systems 
are generally older \citep[][]{Tho05,Nel05}. This phenomenon, known 
as the ``archaeological down-sizing'', appears to be in 
contradiction with the ``hierarchical'' formation scenario, 
but may also indicate that the growth of galaxies in a halo  
does not follow the growth of the halo. 

Recently Gao, Springel \& White ~\citep[2005, see also][]{Wec06,Har06,Jing07,Gao07} 
found that, the half-mass assembly time of a halo is also 
correlated with halo clustering properties on large scales. 
Using $N$-body simulations, these authors find that, 
for halos of a given present mass that is smaller than 
$\Mstar$, the ones that assembled half of their 
final masses earlier are more strongly clustered in 
space. On the contrary, for halos more massive than 
$\Mstar$, the ones that assembled half of their final masses 
later are more strongly clustered. If the star formation history 
is somehow correlated with dark halo formation history, 
as is expected from current theory of galaxy formation, 
these results would indicate that galaxy systems, such as clusters and 
groups, of the same mass but containing different galaxy 
populations should also show different clustering properties.
Observationally, there is evidence to support such connection
\citep[e.g.][]{WY07,Yang06, Bir06}. Although there is still 
discrepancy among the different results, most studies 
show that redder groups are more strongly clustered than 
bluer groups. In particular, the results of~\citet[][]{WY07} suggest
that such color dependence is stronger for groups 
with lower halo masses and becomes insignificant for groups 
with halo masses above $\sim 10^{13}\msunh$. This dependence has 
the same trend as the assembly-time dependence of halo 
clustering, and it is tempting to link these two types 
of dependence. However, as mentioned above, the half-mass 
assembly time may not be a good indicator of the 
typical formation time of stars in a halo. Indeed, using a ``shuffling'' 
technique, \citet{Cro07} found that, the dependence of halo bias
on halo half-mass assembly time can only account for about half of 
the clustering bias seen in red halos in their semi-analytical 
catalogue. Clearly, in order to understand the observational 
results in terms of halo assembly bias, one needs to define 
halo formation times that are more closely related to the formation 
of galaxies in dark matter halos.

The main goal of this paper is to systematically study when various 
characteristic events take place in the halo assembly process and 
how they are correlated with halo mass and with the large-scale 
environments. To this end, we define a number of formation times 
to characterize each halo formation history.    
We study in detail how each of these formation times is 
correlated with halo mass and how the halo 
correlation amplitude depends on these formation times.
Our analysis is based on the``Millennium Simulation''  
~\citep[][]{Sp05}. The paper is organized as follows. 
In Section 2, we briefly describe the simulation and the techniques 
to identify halos and to construct merging trees. In Section 3,
we describe our definitions of halo assembly times and how to estimate 
them from the simulation,  and study how they are correlated
with halo mass. In Section 4, we present the results on the 
assembly-time (age) dependence of halo clustering.  
Finally, in Section 5,  we summarize and discuss the implications of our 
results. 

\section{The Simulation}

In this paper we use the  ``Millennium Simulation'' (MS) carried out by 
the Virgo Consortium~\citep{Sp05}. This simulation follows the evolution
of $2160^3$ dark matter particles in a cubic box of $500\mpch$ on a side.
The particle mass is approximately $8.6\times 10^8\msunh$, which enables
us to study the assembly of halos more massive than around $10^{11}\msunh$ 
with a reasonable mass resolution. The simulation adopts a flat $\Lambda$CDM 
with $\Om=\Odm+\Obaryon=0.205+0.045=0.25$, where $\Odm$ and $\Obaryon$ 
stand for the current densities of dark matter and baryons respectively;
the linear r.m.s. density fluctuation in a sphere of an $8\mpch$ radius, 
$\seight$, equals 0.9; and Hubble expansion parameter $h=0.73$. In total there 
are 63 snapshot outputs between $z=0$ and $z=80$, which are almost evenly 
placed in $\ln(1+z)$ space. In the MS simulation, the characteristic collapsing 
mass, $\Mstar$, defined through $\sigma(\Mstar)=1.69$, is about 
$6\times 10^{12} \msunh$. In order to identify dark halos, the 
Friends-Of-Friends (FOF) algorithm with a linking length $b=0.2$ times 
the mean particle separation is used, so that the structures identified 
(we will call them FOF groups hereafter) have a density
approximately 200 times the mean cosmic density. In addition, by 
smoothing the FOF groups outside-in, each FOF group is also assigned a 
corresponding ``virial halo'' with a ``virial mass'' $M_{\rm v}$,
so that the average density contrast between the ``virial halo'' and 
cosmic critical density $\rho_{\rm c}$, is 
\begin{equation}
\Delta_{\rm v}(z) = 18 \pi^2 + 82 [\Omega_{\rm m}(z)-1] - 39 [\Omega_{\rm m}(z)-1]^2
\end{equation}
\citep{BN98}. The radius at which the density contrast first 
reaches $\Delta_{\rm v}(z)$ defines the virial radius, $R_{\rm v}$, 
of the halo. Despite the fact that $M_{\rm v}$ is slightly 
(typically $5\%$) smaller than the corresponding FOF group mass, there are 
no other significant differences when one studies the accretion history 
of dark halos. In what follows, we always use ``virial halos'' in our 
analysis. Given a cosmological model, we define the virial velocity $\Vh(z)$ 
of a growing halo at redshift $z$ as
\begin{equation}
\label{eq:vcz}
\Vh(z) = \sqrt{G M_{\rm v}(z) \over R_{\rm v}(z)} =
\left[ \frac{\Delta_{\rm v}(z)}{2}\right]^{1/6} \left[M_{\rm v}(z) \, H(z)\right]^{1/3},
\end{equation}
where $R_{\rm v}(z)$ and $H(z)$ are, respectively, the halo virial radius and 
Hubble expansion parameter at redshift $z$. With this, we can follow the growth of the 
virial velocity (which is a measure of the halo gravitational depth) using
the growth of the halo mass.

The merging trees of dark halos in the MS are constructed on the basis of 
sub-halos. In each FOF group, self-bound sub-structures are identified
using {\sc SUBFIND}~\citep{Sp01}. A sub-structure 1 at redshift $z_1$
is considered a progenitor of another sub-structure 2 at $z_2$ ($z_1 > z_2$)
if a certain fraction of its most bound particles are in sub-structure 2.
In addition to sub-halos, each FOF group contains one and only one 
``main virial halo'', with mass $M_{\rm v}$. In our analysis,  
we use $M_{\rm v}$ to construct the mass growth history of the final halo. 
Since in general $M_{\rm v}$ accounts for the central part of an FOF group, 
this treatment naturally avoids the ambiguous case where some accidently 
linked sub-halos that do not belong to the halo also contribute to the 
halo mass.

\section{Halo formation times}
\label{sec:hft}
\begin{figure*}
\centerline{\psfig{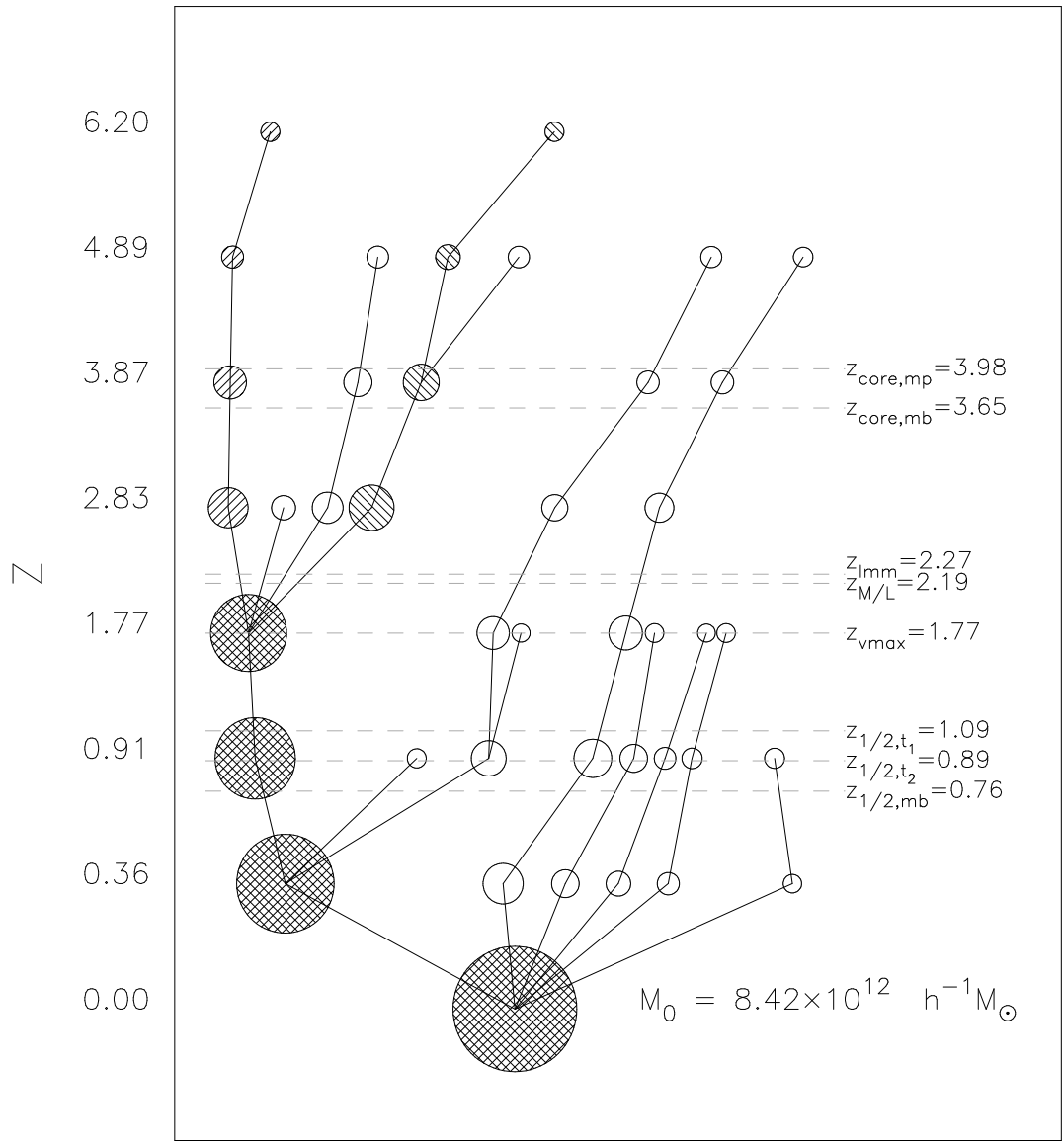}}
\caption{Merging history of a typical MS halo, with all the  defined formation
times marked. Progenitors greater than $4\times 10^{10}\msunh$ are output at
selected redshifts to avoid crowdness. The radius of each circle is
roughly proportional to $M^{1/3}$. Circles filled with hatch lines that are
$45^\circ$ clockwise to the vertical represent the main branch progenitors;
while those filled with hatch lines that are $45^\circ$ counter-clockwise to the vertical
represent the maximum progenitors.}
\label{fig:tree}
\end{figure*}
\begin{figure*}
\centerline{\psfig{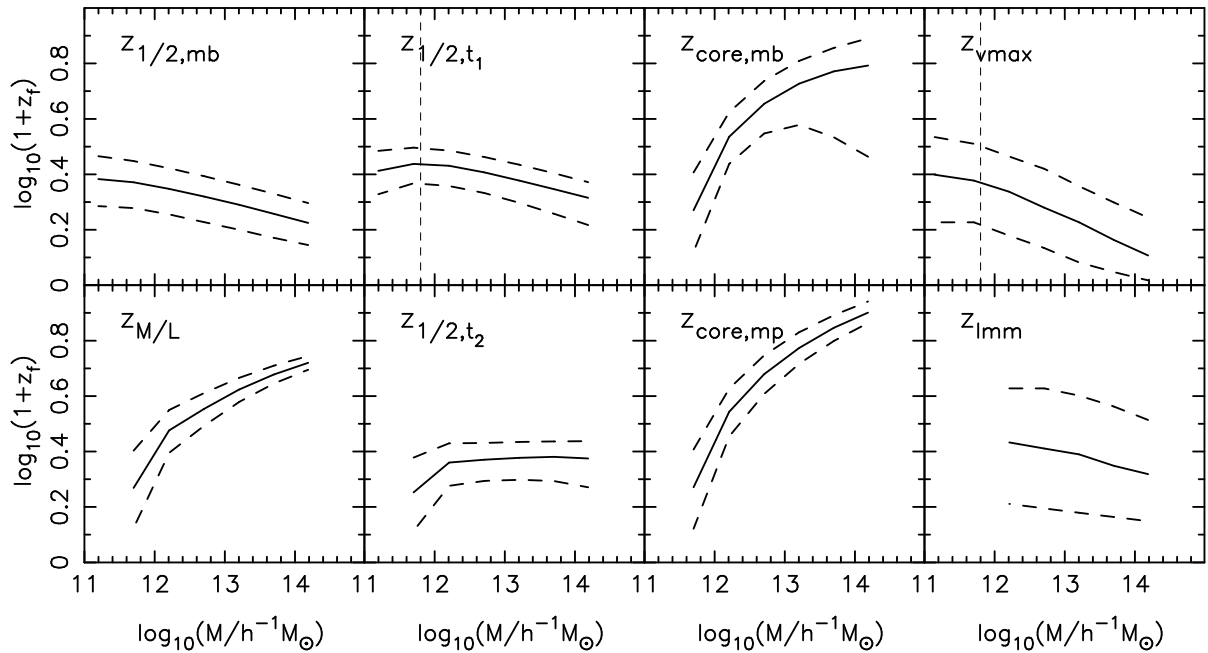}}
\caption{Formation time v.s. halo mass. Solid line represents the
median of each mass bin, while dashed lines represent
20\% and 80\% quantiles, respectively. Note that in the panels
for $\zhalfta$ and $\zvmax$, the slight drop at the low mass end 
(to the left of the vertical dashed line) is due to the finite
mass resolution of the simulation, because small progenitors cannot
all be resolved for halos that are too small.}
\label{fig:mz}
\end{figure*}

As mentioned above, the formation of dark halos is a very complicated
process.  There are two ways to follow the mass growth of a halo with time.
First, one can start with a halo at the present time, pick up the most 
massive progenitor in the {\it adjacent} snapshot at higher redshift, 
and repeat this procedure until the halo mass is so small that it cannot be 
resolved anymore. The mass accretion history of the halo is then represented 
by the growth of the progenitor mass along the ``main branch''. 
Alternatively, one can always look up the most massive progenitor at each 
redshift on the merging tree, and concatenate on these 
progenitors chronologically. The mass accretion history so 
obtained reflects the growth of the ``maximum progenitor''. 
Note that the halo main branch does not always represent the maximum 
progenitor of a halo at each redshift, especially at the early stage of 
a growing halo~\citep[e.g.][]{G04}. Most of the previous studies 
have been concentrated on the ``main branch'' when studying the halo assembly 
histories~\citep[e.g.][]{LC93,Frank02,Gao05,Wec06}. In this paper, 
we use both definitions. We define the following set of parameters 
to characterize the assembly history of a halo:
\begin{enumerate}
\item  $\zhalfmb$:  This is  the redshift  at which the  halo main branch has
  assembled  half of  its final  mass,  $M_{\rm v}(0)$.  This formation  
  time has been widely used in the 
  literature, as mentioned before. 
\item $\zhalfta$: This is the highest redshift at which half of the final halo 
  mass is contained in progenitors with masses ($M_{\rm p}$) greater than 
  $0.02M_{\rm v}(0)$. The same kind of formation time has been 
  used in~\citet[][]{NFW97} to characterize the formation time of 
  a halo and to study how halo concentration is correlated with formation time.  
\item $\zhalftb$: This is the highest redshift at which half of the final halo 
  mass has assembled into progenitors more massive than a fixed mass, 
  $M_{\rm c}=10^{11.5}\msunh$. As shown by~\citet[][]{Frank03} 
  [see also~\citet[][]{Yang03}], halos with masses 
  $\sim M_{\rm c}$ have the minimum 
  mass-to-light ratio, and thus are the most efficient in star formation.
  With $M_{\rm c}=10^{11.5}\msunh$, $\zhalftb$ therefore indicates when star 
  formation starts to prevail the halo assembly history. By definition, 
  only halos more massive than $M_{\rm c}$ have a well-defined $\zhalftb$.
  This formation time is analogous to the formation time, $z_{\rm N06}$, 
  introduced by~\citet{Nei06}. According to~\citet{Nei06}, 
  $z_{\rm N06}$ is the time when the sum of the progenitors above a 
  given minimum mass, reaches half of the present day halo mass.
\item $\zmtl$: This is the redshift when the progenitors more massive than 
  $M_{\rm c}$ have assembled a fraction $f$ of  $M_{\rm v}(0)$. Here the 
  definition of $f$ is based on the non-constant mass-to-light ratio of dark 
  matter halos~\citep{Yang03}. For halos more massive than $M_{\rm c}$, 
  the mass-to-light ratio,  $M_{\rm v}(0)/L$, follows a power law of  
 $M_{\rm v}(0)$, with power-index $\gamma=0.32$~\citep[see][table 1]{Yang03}. 
 We therefore have,
  \begin{equation}
   f = \alpha \frac{L}{ M_{\rm v}(0)} 
    = \frac{1}{2}\left( \frac{ M_{\rm v}(0)}{M_{\rm c}}\right)^{-\gamma},
  \end{equation} 
  where $\alpha$ is a constant, which is set so that $f=\frac{1}{2}$ for
  $M_{\rm v}(0)=M_{\rm c}$.
  Thus, $\zmtl$ essentially reflects the time when a halo becomes capable 
  of forming a fraction of its total stellar mass. Note again that $\zmtl$
  can be defined only for $M_{\rm v}(0)>M_{\rm c}$.
\item $\zcoremb$: This is the highest redshift at which the halo's main branch 
  assembles a mass of $M_{\rm c}$. This formation time therefore indicates 
  when a halo is able to host a relatively bright central galaxy.
\item $\zcoremp$: This is the highest redshift at which the 
  most massive progenitor has reached the mass $M_{\rm c}$.  
  Note that for massive halos, $\zcoremp$ may be different  
  from $\zcoremb$.
\item  $\zvmax$: This  is the  redshift at  which the  halo's virial
  velocity $\Vh(z)$  reaches its maximum value  over the entire mass accretion 
  history.  According to equation~(\ref{eq:vcz}), 
  the value of $\Vh(z)$  is expected 
  to increase  (decrease) with time, if the time scale  for mass accretion is 
  shorter (longer) than the  time  scale of  the  Hubble  expansion.  
  Therefore,  $\zvmax$ indicates the time when the halo mass accretion transits
  from a fast accretion phase to a slow accretion phase~\citep{Zhao03,Li07}.
\item $\zlmm$: Last major merger time. Here we define a major merger
   as the event when the mass ratio between the smaller halo and the main 
   halo is no less than 1/3. The last major merger time is defined 
   to be the one when the last major merger occurred on the main branch 
   of an assembling halo.
\end{enumerate}

Once the merging history of a halo is given, it is quite 
straightforward to determine the formation times defined above.
The only exception is $\zlmm$. Since the mass transfer from
the merging halo to the main halo is a gradual process,
a merger in general takes several snapshots
to complete. Thus, if we used the halo mass increase  in one time 
step, we would find only a small number of events in which 
the increase in the halo mass in a time step is large enough to be 
qualified as major mergers. In order to circumvent this problem, 
we start from one snapshot, and trace the progenitors 
(including those of sub-halos) back to all the snapshots within 
a 1-Gyr interval.  As long as there is a progenitor with 
mass exceeding 1/3 of the main-branch halo mass at the same 
time, a major merger event is identified. The choice of 1 Gyr 
is not crucial; our tests using 0.5 Gyr or 1.5 Gyr give almost
the same results.

As illustration, we plot in Fig.~\ref{fig:tree} an actual merging 
history of a typical halo selected from the MS simulation, 
with all formation redshifts defined above marked. As one can 
see, the different definitions give very different values 
of the formation redshift, and they capture quite different 
aspects of the assembly history of a dark matter halo.  

In Fig.~\ref{fig:mz} we show each of the formation redshifts 
as a function of halo mass. In each panel, the solid line 
represents the median in each mass bin, while the dashed lines 
represent 20\% and 80\% percentiles, respectively. 
As one can see, less massive halos generally have higher values of 
$\zhalfmb$, $\zhalfta$, $\zvmax$ and $\zlmm$ than massive ones, 
i.e. these formation redshifts have a negative correlation with 
halo mass. Since these formation times are defined in a 
self-similar manner, i.e., do not involve any particular mass 
scale, it is not surprising that they show a similar 
``bottom-up'' trend, a consequence of hierarchical clustering.
Nevertheless, $\zhalfmb$, $\zhalfta$, $\zvmax$ and $\zlmm$ still 
represent quite different epoches of halo formation history, 
which can be seen from their different values and scatter. 
For all halo masses, both $\zvmax$ and $\zlmm$ have 
scatter that is much larger than $\zhalfmb$, $\zhalfta$.
This indicates that both $\zvmax$ and $\zlmm$ are more 
sensitive to the details of the halo assembly history. 

On the other hand,  the other four formation redshifts, 
$\zmtl$, $\zhalftb$, $\zcoremb$ and $\zcoremp$, 
all show positive correlation with the halo mass, in the sense 
that more massive halos experience these events earlier.
For massive halos, $\zcoremb$ is lower than $\zcoremp$ and has 
larger scatter, which is due to the fact that for some 
massive halos, the most massive progenitors are not in 
the main branch. The trend is particularly strong for $\zmtl$, $\zcoremp$
and $\zcoremb$. A halo of $10^{12}h^{-1}{\rm M}_\odot$ 
assembles a progenitor of mass $10^{11.5}h^{-1}{\rm M}_\odot$ 
typically at $z\sim 1$, while such a progenitor forms at $z\sim 5$  
for halos with masses $\ga 10^{14}h^{-1}{\rm M}_\odot$. 
Since $\zmtl$, $\zcoremb$ and $\zcoremp$ are the redshifts 
that characterize when a halo was able to host a 
relatively bright galaxy, the results shown here suggest that 
massive galaxies can form much earlier in massive halos 
than in low-mass halos. If star formation in these massive 
galaxies was eventually quenched as their stellar masses
reach to some value, as is the case in the current AGN 
feedback model, or as they merge into a massive 
halo where radiative cooling becomes inefficient
\citep[e.g.][]{Churazov05,Cro06,Cat08}, one would expect 
that the star formation activity shifts with the passage of time 
from high-mass systems to the low-density field. This may be 
related to the observed ``down-sizing'' effect that massive 
galaxies in present-day clusters in general have old stellar 
populations with little star formation activities, and 
most star formation activities at the present time 
have shifted to low-mass systems. This shift is perfectly consistent 
with the hierarchical formation of dark matter halos, provided that there 
are some mechanisms that can quench star formation in massive galaxies.  
As we have shown, more massive halos indeed assemble their 
masses later, but the formation of massive galaxies can actually 
start earlier in their progenitors. 

As mentioned before, the formation redshift $\zhalftb$ defined 
here  is similar to the formation time $z_{\rm N06}$ introduced
by~\citet{Nei06}. However, the halo mass-dependence we obtain 
here is quantitatively different from theirs. At the massive 
end, the results of \citet{Nei06} show continuous increase 
of the formation redshift with halo mass, while ours show 
a flattened relation. Note that \citet{Nei06} used the extended 
Press-Schechter formalism to generate halo merging trees,
while we obtain halo merging trees directly  from $N$-body 
simulation. We suspect that the discrepancy may be due to the 
inaccuracy of the extended Press-Schechter formalism. 

\begin{figure*}
\centerline{\psfig{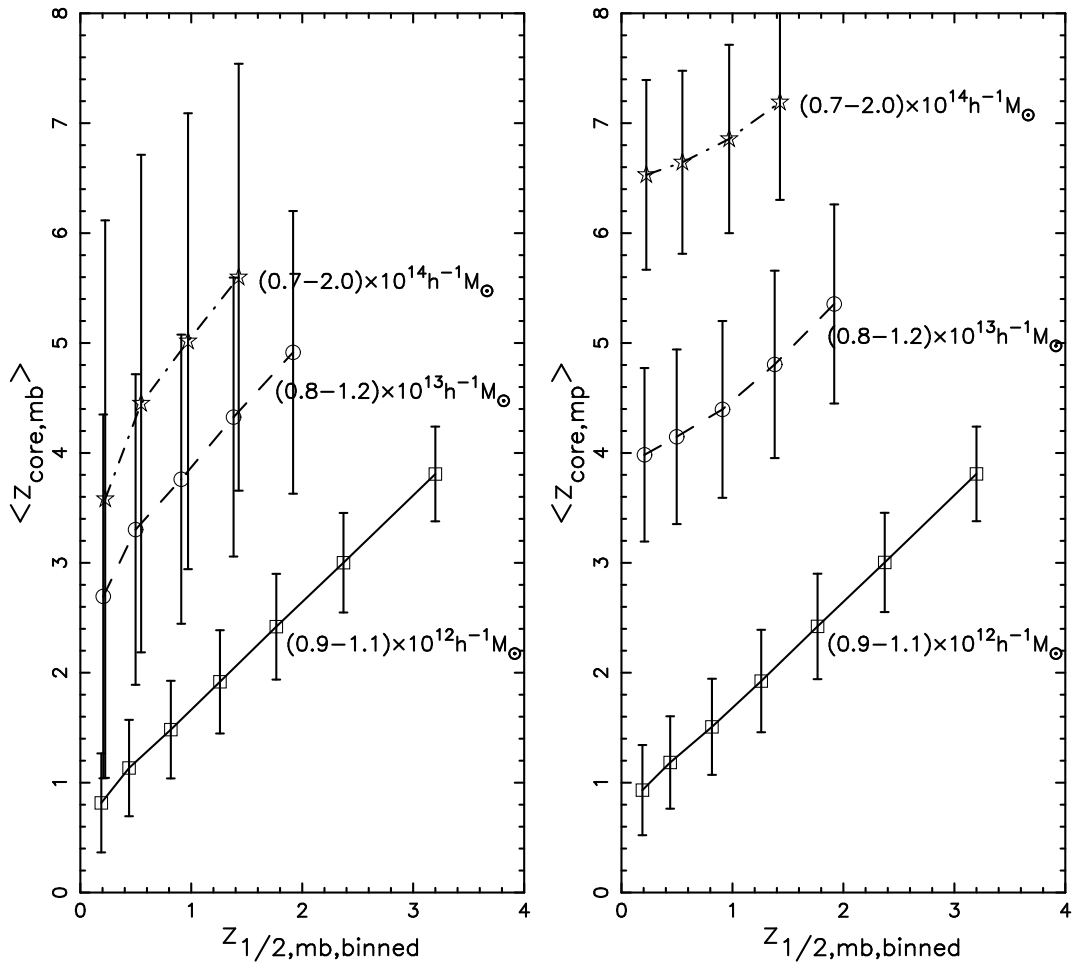}}
\caption{Compare $\zcoremb$ and $\zcoremp$ with $\zhalfmb$. Each line
represents the result for halos with the mass marked aside. For each point,
horizontal axis is the binned $\zhalfmb$, and vertical axis is the average
$\zcoremb$ or $\zcoremp$ in the bin; error bars represent the standard 
deviation.
More massive halos generally have lower $\zhalfmb$ but
higher $\zcoremb$ or $\zcoremp$. Error bars in the right panel are
generally smaller than in the left panel,
indicating that the maximum progenitor could substantially deviate
from the main branch, especially for massive halos at early time.}
\label{fig:zhalfzcore}
\end{figure*}

In Fig.~\ref{fig:zhalfzcore} we show $\zcoremb$ and  
$\zcoremp$ versus $\zhalfmb$ for halos of different masses. 
For low-mass halos, $M_{\rm v}\sim 10^{12}h^{-1}{\rm M}_\odot$, 
$\zcoremb$ and $\zcoremp$ are very similar to  $\zhalfmb$. 
However, for halos more massive than $10^{13}\msunh$, 
$\zcoremb$ and $\zcoremp$ are both higher than $\zhalfmb$. 
In particular, for halos with $M_{\rm v}\ga 10^{14}h^{-1}{\rm M}_\odot$, 
$\zcoremp\sim 7$, without depending strongly on the half-mass 
formation redshift, $\zhalfmb$. This shows again that, 
for massive halos, the progenitors that can host massive
galaxies can form much earlier than when the halos assemble 
most of their masses. Thus, although dark halos form 
hierarchically, star formation may appear ``anti-hierarchical''
at late epochs when many halos in which star formation was 
efficient have merged into massive systems.          

\section {The Dependence of Halo Clustering on Formation Times}

Halos are biased tracer of the dark matter distribution.
On large scales the auto-correlation function of dark halos,
$\xihh$, is expected to be parallel to that of the mass, $\ximm$,
so that one can write
\begin{equation}\label{linear_bias}
\xihh(M,r)=b^2\ximm(r),
\end{equation}
where $b$ is the so-called halo bias factor. Analytical models and 
$N$-body simulations have shown that the halo bias factor depends
strongly on halo mass. Halos more massive than $\Mstar$ are more 
strongly clustered than the underlying mass (i.e. $b>1$),  
while low-mass halos are less clustered than the mass (i.e. $b<1$) 
\citep[e.g.][]{Mo96,Jing98,ST99,SMT01}.
More recently, using $N$-body simulations,  \citet{Gao05} found that 
for halos with a fixed mass, the halo bias factor $b$ also depends on 
the time when the halo first assembles half of its mass, i.e. on 
$\zhalfmb$, in the sense that halos with higher $\zhalfmb$ are more 
strongly clustered (i.e. have a higher bias factor). This assembly bias 
is found to be stronger for halos of lower mass. Subsequent 
investigations using different simulations have confirmed this result
~\citep[e.g.][]{Wec06,Har06,Zhu06,Jing07}, and theoretical 
models have been proposed to understand the origin of such assembly 
bias \citep[e.g.][]{Wang07,Sand07,Hahn07,Des07,KN07}. In most of these
earlier investigations, the assembly bias is analyzed in terms of 
the half-mass assembly time, $\zhalfmb$. However, as we discussed above,
although $\zhalfmb$ may be a good quantity to characterize the formation 
of the main body of a halo, it does not characterize other aspects 
of the halo formation histories that may be more closely related to the 
formation of galaxies in halos. With the various formation times
we have obtained in this paper, it is interesting to investigate how 
the clustering of halos depends on these different formation times.

\begin{figure*}
\centerline{\psfig{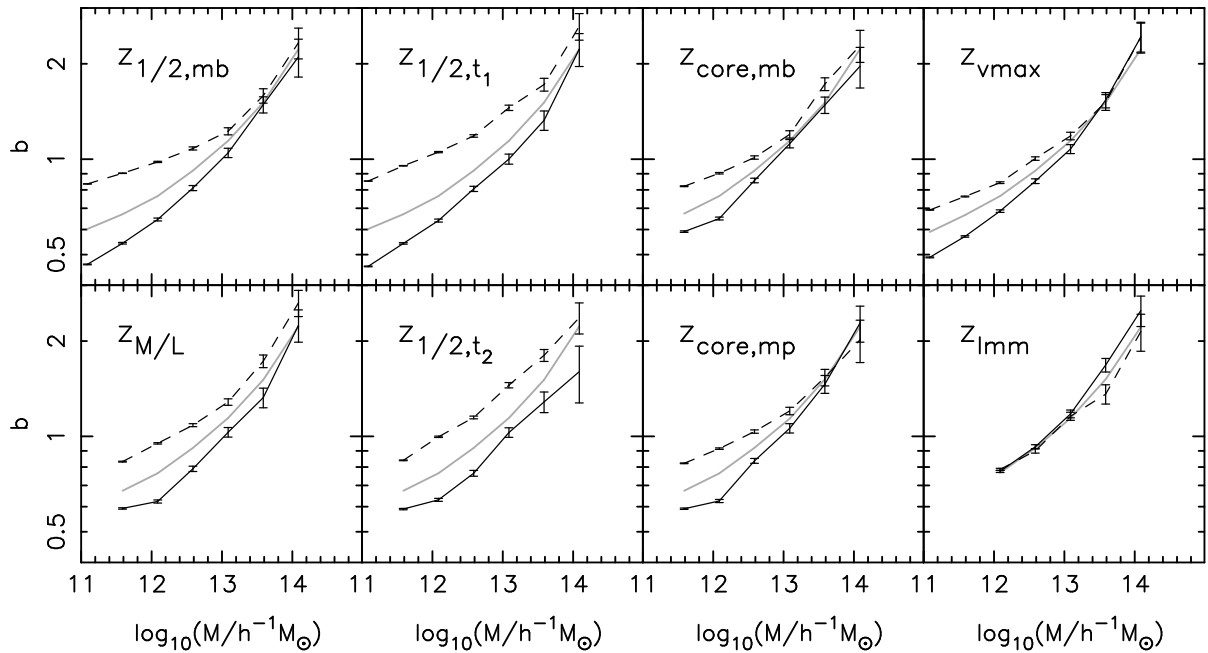}}
\caption{Age dependence of halo bias. Formation time used is indicated
in each panel. Dashed lines are for oldest $20\%$ halos while solid lines
are for youngest $20\%$ halos; the thick gray lines represent the bias of 
all the halos regardless of their ages. Error barshows the Poisson error.}
\label{fig:MSbias}
\end{figure*}
In order to study the formation-time dependence of halo clustering, 
we use a 3-dimensional Fast-Fourier-Transform to derive the 
two-point correlation function of dark matter halos as well 
as dark matter particles on large scales. We then estimate the halo bias 
$b$ for a given halo mass, using the square root of
the ratio of the two-point correlation function of halos and that of 
dark matter, averaged over data points in $8\mpch\le r< 20\mpch$.
This interval of $r$ is chosen to ensure that the clustering is 
still in the quasi-linear regime where the linear bias relation  
(\ref{linear_bias}) is a good approximation. In  Fig.~\ref{fig:MSbias} 
we show how halo bias depends on the formation times we have defined
for halos of different masses. The dashed line in each panel 
shows the bias factor of the oldest 20\% population among all 
halos as a function of halo mass, while the solid line shows the 
corresponding result of the youngest 20\%. For comparison, we also 
show, as the grey line, the results for the total population
without separation according to formation time. The errorbar
on each data point is estimated  through the error propagation function 
based on the Poisson noise of each data point of the halo 
two-point correlation in $8\mpch\le r< 20\mpch$.

As one can see, the bias factor of the total population increases 
with halo mass, and the increase is more rapid at the massive end.
This result of mass-dependence of the halo bias factor is in good 
agreement with the results obtained in earlier investigations. 
In addition, for halos of a given mass, the bias factor also 
depends on the formation redshifts, although the strength 
of the dependence is not always the same for different definitions.
The result based on $\zhalfmb$ is very similar to that obtained by 
\citet{Gao05}, even though the result here is based on ``virial'' 
halos while Gao \etal ~used FOF groups. 
With the exception of the case with $\zlmm$, where no significant
age dependence is found for any halo masses, the strength of 
the age dependence  in general decreases with increasing halo mass.  
For halos more massive than  $10\Mstar$, we do not see any 
significant difference between halos of different ages. 
However, the noise here is too large due to the small number of systems 
available from the simulation, and so our data is not able to 
reveal the weak reversed trend, namely that the youngest halos 
are more strongly clustered, at the very massive end seen in 
some simulations \citep[e.g.][]{Jing07,Wet07}. The strongest age dependence 
is seen in the cases of $\zhalfmb$, $\zhalfta$ and $\zhalftb$. 
Note that all these three formation redshifts are based on the 
properties of half of the final halo masses. This suggests that 
the assembly of the main parts of halos, especially 
the low-mass ones, may be affected significantly by large-scale 
environments. On the other hand, for the definitions 
that are based on the formation of a progenitor of a fixed mass, 
such as $\zcoremb$ and $\zcoremp$, the age dependence is weaker, 
particularly for halos with masses much higher than the 
progenitor mass, $M_{\rm c}$, used in the definition. As shown earlier,
such progenitors in massive halos usually form at high redshifts
where the large-scale environmental effects may not yet have time 
to develop. The age dependence based on $\zvmax$ is also weaker
than that based on $\zhalfmb$, presumably because the halo density
involved in defining $\zvmax$ is relatively high and so the 
the mass assembly before $\zvmax$ is less affected by the 
large-scale environment than that before $\zhalfmb$. 
Fig.~\ref{fig:MSbias} also shows that there is almost no dependence of 
the bias factor on $\zlmm$. This suggests that major mergers 
may be controlled by the properties of the local density field, without 
being strongly modulated by large-scale environments. 
This is consistent with the result of ~\citet[][]{Per03}, who found that,
for halos with very recent major mergers (within the past $10^8$ years), there is
no difference in the halo bias compared with all the halos of similar mass.

The formation-time dependence of the halo bias presented above 
may have important implications. Previous studies suggest that 
the halo assembly bias may introduce observable effects in  
the large-scale clustering of galaxies \citep[e.g.][]{Nei06,Cro07}.
Using a large group catalogue constructed from the SDSS, \citet{WY07} 
found that groups with a redder central galaxy or a redder average 
color of member galaxies show stronger clustering  
\citep[see also][]{Yang06}. 
Because of the complexity of halo assembly, it is unclear 
which aspects of the halo formation history are more closely 
related to the colors of the galaxies that form in halos.
By ``shuffling'' galaxies contained in halos of similar mass 
or formation time $\zhalfmb$, \citet{Cro07} found, in their 
semi-analytical model, that the $\zhalfmb$-dependence of halo 
clustering  can account at most half of the clustering bias of red 
galaxies. This implies that the difference in $\zhalfmb$ alone 
may not be sufficient to account for the colors of galaxies.
This result is not surprising, because the assembly history 
of a halo is quite complicated and it is not expected that 
$\zhalfmb$ can provide a full characterization of such history. 
As demonstrated above, each of the eight formation times defined 
in this paper catches a different aspect of the halo formation 
history. It would be interesting to if see some combinations 
of these formation times are better correlated with the 
properties of galaxies. As we have shown, the assembly bias 
becomes insignificant for halos more massive than 
$10^{13}h^{-1}\msun$ for most of the definitions of the 
assembly times. This may be the reason why the color-dependence  
of galaxy group clustering is significant only for groups
less massive than $\sim 10^{13}h^{-1}\msun$~\citep{WY07}.

Our results also show that there is virtually no dependence
of halo bias on $\zlmm$, the redshift of last major merger.  
In the literature, it has been suggested that major merger 
may effectively shut off the star formation in a galaxy
~\citep[e.g.,][]{Hq89,MH96,Sp05b,Kang06}, 
and hence $\zlmm$ should be correlated with the current color 
of the central galaxy. However, if major mergers were the 
main reason to make a galaxy red, there would be no 
color-dependence of the clustering amplitude of galaxy groups, 
contrary to the observational results of \citet{WY07}.
This suggests that major mergers alone cannot explain the 
red color of central galaxies. It is possible that the reddening 
of a central galaxy is the  accumulative effect of a series of 
events triggered by, say, minor mergers, rather than a dominant 
major merger \citep[e.g.][]{Geo08}.

\section{Discussion and Conclusions}
\label{sec:concl}

In this paper we examine the complexity of dark halo mass 
assembly history using the MS and using different formation times to 
characterize the various aspects of halo formation histories.  
We find that, formation times defined according to the assembly 
of a fixed fraction of the halo final mass characterize the 
hierarchical clustering, in the sense that halos of higher masses 
on average have later formation time. 
On the other hand, formation times defined by the formation 
of progenitors of a fixed mass where star formation is expected 
to be efficient, clearly show ``anti-hierarchical'' behavior, in the 
sense that halos of higher masses have earlier formation time. 
If some feedback processes can terminate the star 
formation in these progenitors, we would expect that galaxies 
in massive halos are redder, consistent with observation.  
We would also expect that the star-formation activities 
should shift with time from high-mass to low-mass halos,
and so the observed ``down-sizing'' in star formation is not in 
conflict with hierarchical clustering.  

 We also study how the clustering of dark matter halos 
depends on the various formation times defined in this paper. 
We find that halo bias shows a strong positive correlation 
with halo mass, in good agreement with earlier results.
For fixed halo mass, our results confirm a positive correlation
between halo formation time, $\zhalfmb$, and halo clustering strength
\citep[e.g.][]{Gao05,Wec06}. 
The strength of this dependence increases with decreasing 
halo mass. For halos more massive than $10^{14}\msunh$, we do not 
find a clear reversal of the assembly bias. 
In general, for halos less massive than $10^{14}\msunh$, 
there is a positive correlation between the various formation 
times defined and halo clustering strength, with the correlation 
being stronger for lower halo masses. However, the correlation 
amplitude is quite different when different formation time is 
considered. The strongest age dependence of halo clustering 
is seen on $\zhalfta$ and $\zhalftb$. There is virtually no age 
dependence of halo clustering on halo last major merger time, 
$\zlmm$, and the dependence on $\zmtl$, $\zcoremb$, $\zcoremp$ is 
moderate. If the typical age of stars in a halo is correlated 
with halo assembly history in some way, then halos with fixed 
mass but containing redder member galaxies are expected to be
more strongly clustered, and this color-dependence is expected 
to be weaker for more massive systems. This is consistent with 
recent observations. However, since there is virtually 
no dependence of halo clustering on $\zlmm$, the typical 
color of galaxies in a halo is not expected to be determined 
by the last major merger time of its host halo.  

\section*{Acknowledgments}
HJM would like to acknowledge the support of NSF AST-0607535, 
NASA AISR-126270 and NSF IIS-0611948.
The {\it Millennium Simulation} was carried out as part of the
programme of the Virgo Consortium on the Regatta supercomputer of the
Computing Centre of the Max-Planck Society in Garching.

\label{lastpage}

\end{document}

%% file: macros_yun.tex
\newcommand{\Vh}{V_{\rm vir}}

\newcommand{\zhalfmb}{z_{\rm 1/2,mb}}
\newcommand{\zhalfta}{z_{\rm 1/2,t_1}}
\newcommand{\zhalftb}{z_{\rm 1/2,t_2}}
\newcommand{\zmtl}{z_{\rm M/L}}
\newcommand{\zvmax}{z_{\rm vmax}}
\newcommand{\zcoremb}{z_{\rm core,mb}}
\newcommand{\zcoremp}{z_{\rm core,mp}}
\newcommand{\zlmm}{z_{\rm lmm}}

\newcommand{\etal}{{et al.}}

\newcommand{\msun}{\>{\rm M_{\odot}}}

\newcommand{\beq}{\begin{equation}}
\newcommand{\eeq}{\end{equation}}

\newcommand{\mpch}{\>h^{-1}{\rm {Mpc}}}

\newcommand{\Mstar}{M_*}

\newcommand{\Om}{\Omega_{\rm M}}
\newcommand{\Odm}{\Omega_{\rm dm}}
\newcommand{\Obaryon}{{\Omega_{\rm b}}}

\newcommand{\seight}{\sigma_8}

\newcommand{\msunh}{\>h^{-1}\rm M_\odot}

\newcommand{\xihh}{\xi_{\rm hh}}
\newcommand{\ximm}{\xi_{\rm mm}}

\newcommand{\apj}{ApJ}

\newcommand{\apjl}{ApJL}

\newcommand{\mnras}{MNRAS}

\newdimen\hssize
\newdimen\hdsize 
\def\ltsima{$\; \buildrel < \over \sim \;$}
\def\lsim{\lower.5ex\hbox{\ltsima}}
\def\gtsima{$\; \buildrel > \over \sim \;$}
\def\gsim{\lower.5ex\hbox{\gtsima}}